\documentclass[aps,prl,twocolumn,showpacs,superscriptaddress]{revtex4-1}  
\usepackage{graphicx} 
\usepackage{bm}       
\usepackage{amssymb}  
\usepackage[bookmarks=true,dvipdfm]{hyperref}
\usepackage{color}

  \hypersetup{
   colorlinks=true,
   pdfborder={0 0 0},
   citecolor=blue,
   linkcolor=blue,
   urlcolor=blue
 }

\begin{document}
\author{A.~Lemasson} 
\affiliation{National Superconducting Cyclotron Laboratory, Michigan State University, East Lansing, Michigan 48824, USA}
\author{H.~Iwasaki} 
\affiliation{National Superconducting Cyclotron Laboratory, Michigan State University, East Lansing, Michigan 48824, USA}
\affiliation{Department of Physics and Astronomy, Michigan State University, East Lansing, Michigan 48824, USA}
\author{C.~Morse} 
\affiliation{National Superconducting Cyclotron Laboratory, Michigan State University, East Lansing, Michigan 48824, USA}
\affiliation{Department of Physics and Astronomy, Michigan State University, East Lansing, Michigan 48824, USA}
\author{D.~Bazin} 
\affiliation{National Superconducting Cyclotron Laboratory, Michigan State University, East Lansing, Michigan 48824, USA}
\author{T.~Baugher} 
\affiliation{National Superconducting Cyclotron Laboratory, Michigan State University, East Lansing, Michigan 48824, USA}
\affiliation{Department of Physics and Astronomy, Michigan State University, East Lansing, Michigan 48824, USA}
\author{J.S.~Berryman} 
\affiliation{National Superconducting Cyclotron Laboratory, Michigan State University, East Lansing, Michigan 48824, USA}
\author{A.~Dewald}
\affiliation{Institut f{\"u}r Kernphysik der Universit{\"a}t zu K{\"o}ln, D-50937 K{\"o}ln, Germany}
\author{C.~Fransen}
\affiliation{Institut f{\"u}r Kernphysik der Universit{\"a}t zu K{\"o}ln, D-50937 K{\"o}ln, Germany}
\author{A.~Gade}
\affiliation{National Superconducting Cyclotron Laboratory, Michigan State University, East Lansing, Michigan 48824, USA}
\affiliation{Department of Physics and Astronomy, Michigan State University, East Lansing, Michigan 48824, USA}
\author{S.~McDaniel}
\affiliation{National Superconducting Cyclotron Laboratory, Michigan State University, East Lansing, Michigan 48824, USA}
\affiliation{Department of Physics and Astronomy, Michigan State University, East Lansing, Michigan 48824, USA}
\author{A.~Nichols}
\affiliation{Department of Physics, University of York, Heslington, York YO10 5DD, United Kingdom}
\author{A.~Ratkiewicz}
\affiliation{National Superconducting Cyclotron Laboratory, Michigan State University, East Lansing, Michigan 48824, USA}
\affiliation{Department of Physics and Astronomy, Michigan State University, East Lansing, Michigan 48824, USA}
\author{S.~Stroberg}
\affiliation{National Superconducting Cyclotron Laboratory, Michigan State University, East Lansing, Michigan 48824, USA}
\affiliation{Department of Physics and Astronomy, Michigan State University, East Lansing, Michigan 48824, USA}
\author{P.~Voss}
\affiliation{National Superconducting Cyclotron Laboratory, Michigan State University, East Lansing, Michigan 48824, USA}
\affiliation{Department of Physics and Astronomy, Michigan State University, East Lansing, Michigan 48824, USA}
\affiliation{Simon Fraser University, Burnaby, British Columbia, V5A 1S6 Canada}
\author{R.~Wadsworth}
\affiliation{Department of Physics, University of York, Heslington, York YO10 5DD, United Kingdom}
\author{D.~Weisshaar}
\affiliation{National Superconducting Cyclotron Laboratory, Michigan State University, East Lansing, Michigan 48824, USA}
\author{K.~Wimmer}
\affiliation{National Superconducting Cyclotron Laboratory, Michigan State University, East Lansing, Michigan 48824, USA}
\author{R.~Winkler}
\affiliation{National Superconducting Cyclotron Laboratory, Michigan State University, East Lansing, Michigan 48824, USA}
\title{   Observation    of   mutually   enhanced    collectivity   in
  self-conjugate~$^{76}_{38}$Sr$_{38}$ }
\date{\today}
\begin{abstract}
The  lifetimes of the  first 2$^{+}$  states in  the neutron-deficient
$^{76,78}$Sr isotopes were measured  using a unique combination of the
$\gamma$-ray line-shape method and two-step nucleon exchange reactions
at intermediate energies.  The transition rates for the 2$^{+}$ states
were                  determined                 to                 be
$B$(E2;2$^{+}$$\to$0$^{+}$)~=~2220(270)~e$^{2}$fm$^{4}$~for~$^{76}$Sr
and  1800(250)~e$^{2}$fm$^{4}$~for~$^{78}$Sr,  corresponding to  large
deformation       of       $\beta_2$~=~0.45(3)~for~$^{76}$Sr       and
0.40(3)~for~$^{78}$Sr.  
The present data provide experimental evidence for mutually enhanced
collectivity that occurs at $N$~=~$Z$~=~38.
The systematic behavior of  the excitation energies and $B$(E2) values
indicates   a   signature   of   shape   coexistence   in   $^{76}$Sr,
characterizing  $^{76}$Sr  as one  of  most  deformed  nuclei with  an
unusually reduced $E$(4$^{+}$)/$E$(2$^{+}$) ratio.
\end{abstract}

\maketitle 
Deformation  of   finite  quantum   systems  is  a   manifestation  of
spontaneous symmetry breaking.
In    analogy    to    the    Jahn-Teller    effect    in    molecular
physics~\cite{Jah37}, the  coupling between collective  vibrations and
degenerate excitations of individual  nucleons plays an important role
in inducing nuclear ground-state deformation~\cite{Boh76}.
In nuclei, pairing correlations  can also compete with the deformation
driving particle-vibration coupling, further highlighting rich aspects
of this many-body quantum system.
Self-conjugate nuclei have provided challenges to our understanding of
the  role of deformation  driving mechanisms  including neutron-proton
correlations~\cite{Lis82,Mac00,*Mac01}.
In  $N$~=~$Z$  nuclei,  proton  and  neutron  shell  effects  can  act
coherently, promoting an extreme  sensitivity of nuclear properties to
small  changes of  nucleon  numbers.  
A  well known  region is  the  middle of  the $pfg$  shell, where  the
nuclear      shape     evolves      drastically      from     triaxial
($^{64}$Ge~\cite{Sta07}), to transitional ($^{68}$Se \cite{Obe09}), to
oblate  ($^{72}$Kr \cite{Gad05})  shapes  with a  gradual increase  of
collectivity 
accompanied by the  intrusion of the  deformation driving 
$g_{9/2}$ orbital~\cite{Naz85,Lan03,Has07}.
This region  at $N$~=~$Z$  represents a unique  location in  the nuclear
chart,  where a strong  enhancement of  collectivity is  expected from
sizable  numbers of valence  protons and  neutrons occupying  the same
orbitals.
However,
questions  remain to  be answered  regarding the  magnitude and  location of  
--  and evolution towards -- the maximum collectivity.
In this Rapid Communication, we report on the first measurement of the
lifetime  of  the first  2$^+$  state  in  the self-conjugate  nucleus
$^{76}$Sr at $N$~=~$Z$~=~38.
We deduce       the      reduced      transition      probability
$B$(E2;2$^{+}$$\to$0$^{+}$)  (noted  $B$(E$2$$\downarrow$)  hereafter).
This quantity provides a direct measure of quadrupole collectivity and
often  serves as  a good  indicator of  the  ground-state deformation,
particularly for well-deformed nuclei.
A  lifetime  measurement  was   also   performed  for  $^{78}$Sr  as  a
reference.
Of  particular   interest  are   the  very  low   excitation  energies
$E$(2$^{+}$)    of   the   first    2$^{+}$   states    measured   for
$^{76}$Sr~\cite{Lis82}  and   neighboring  $^{78}$Sr~\cite{Lis82}  and
$^{80}$Zr~\cite{Lis82},   which  suggest   the   occurrence  of   large
deformation   at   $A$~$\sim$~80   in   agreement   with   theoretical
predictions~\cite{Sar09,Naz85,Bon85,Lal95,Pet02,Mar92,Lan03,Has07,Del10}.
For $^{76}$Sr, the Gamow-Teller  strength distribution measured in the
$\beta$-decay      study      strongly      favors      a      prolate
deformation~\cite{Nac04}.
Rotational properties of the yrast band are well established in medium
and high-spin states up to 22$^{+}$ in $^{76}$Sr~\cite{Dav07}.
However,          the          measured          energy          ratio
$R_{4/2}$~=~$E$(4$^{+}$)/$E$(2$^{+}$)   of  2.85   could   indicate  a
triaxial   deformation~\cite{Lis90}  with   a   reduced  collectivity,
hampering the establishment of a consistent picture for $^{76}$Sr.
Here,  based  on new  $B$(E$2$$\downarrow$)  data,  we  provide a  new
perspective  on the  evolution  of collectivity  at  $N$~=~$Z$ and  an
insight  into  the  character   of  the  ground-state  deformation  of
$^{76}$Sr.

Exploring  deformation  of  heavy  $N$~$=$~$Z$  nuclei  represents  an
experimental challenge.
In this work, the $\gamma$-ray line-shape method~\cite{Ter08,Doo10}
was applied to measure  the  lifetimes  of  the 2$^{+}$  states  of
$^{76,78}$Sr, which were produced in two-step nucleon exchange reactions 
at intermediate energies~\cite{Gad09}.
A dedicated  configuration for  the present lifetime  measurements was
employed  for  the first  time  using  the  Segmented Germanium  Array
(SeGA)~\cite{Mue01}   at   the   National  Superconducting   Cyclotron
Laboratory (NSCL).
Also, the  present reaction  scheme allows access  to nuclei  that are
neutron-deficient  but with higher~$Z$  than available  primary beams,
providing an  attractive alternative to the use  of fusion evaporation
or fragmentation reactions in lifetime measurements.
This approach facilitates a  clear identification of reaction products
and a sizable population of excited states simultaneously.

\begin{figure}[t]
\begin{center}
  \includegraphics[width=1\columnwidth]{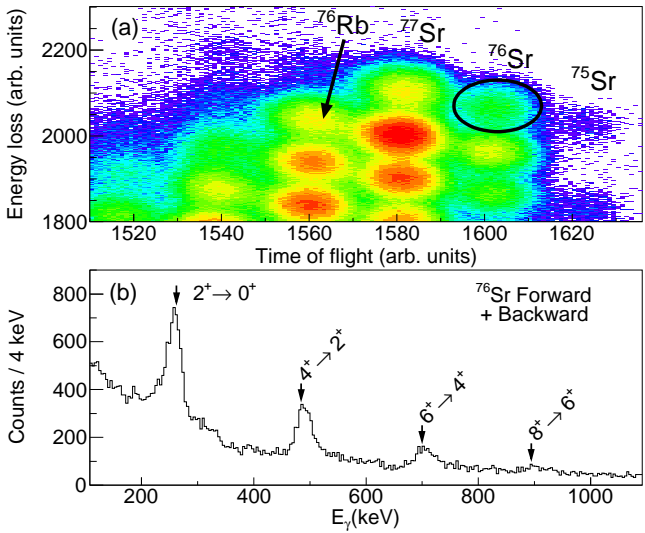}
 \caption{\label{fig:fig1} (color  online) (a) Particle identification
   of reaction residues for  the S800 setting optimized for $^{76}$Sr.
   (b)  Doppler-shift corrected  $\gamma$-ray  spectrum for  $^{76}$Sr
   (with   $\beta_{\text{mid}}=0.396$).   The   arrows  indicate   the  known
   $\gamma$-ray energies~\cite{Lis90}.   
}
\end{center}
\end{figure}

The experiment was performed at the Coupled Cyclotron Facility of NSCL
at Michigan State University.
Secondary  beams  of  $^{76}_{37}$Rb$_{39}$ and  $^{78}_{37}$Rb$_{41}$
were produced by reactions  of a primary beam of $^{78}_{36}$Kr$_{42}$
at 140~MeV/nucleon  with a  $^9$Be target and  separated by  the A1900
fragment separator~\cite{Mor03} using an Al degrader.
The momentum acceptance of the A1900  was set to 0.5\%.
The   resultant  beams   were  $\approx$30$\%$   $^{76}$Rb   at  104.5
MeV/nucleon  and $\approx$70$\%$  $^{78}$Rb at  101.6  MeV/nucleon for
each setting.
The available intensity was typically around 4$\times$10$^{4}$ pps for
$^{76}$Rb,  while  for  $^{78}$Rb the  beam  was  used  at a  rate  of
1$\times$10$^{5}$~pps.
The  $^{76}_{38}$Sr$_{38}$  and  $^{78}_{38}$Sr$_{40}$  isotopes  were
produced and  studied using  the secondary nucleon  exchange reactions
$^9$Be($^{76}$Rb,$^{76}$Sr$\gamma$)X                                and
$^9$Be($^{78}$Rb,$^{78}$Sr$\gamma$)X,      respectively,      on     a
376-mg/cm$^2$-thick $^9$Be reaction target. 
\begin{figure}
\begin{center}
  \includegraphics[width=\columnwidth]{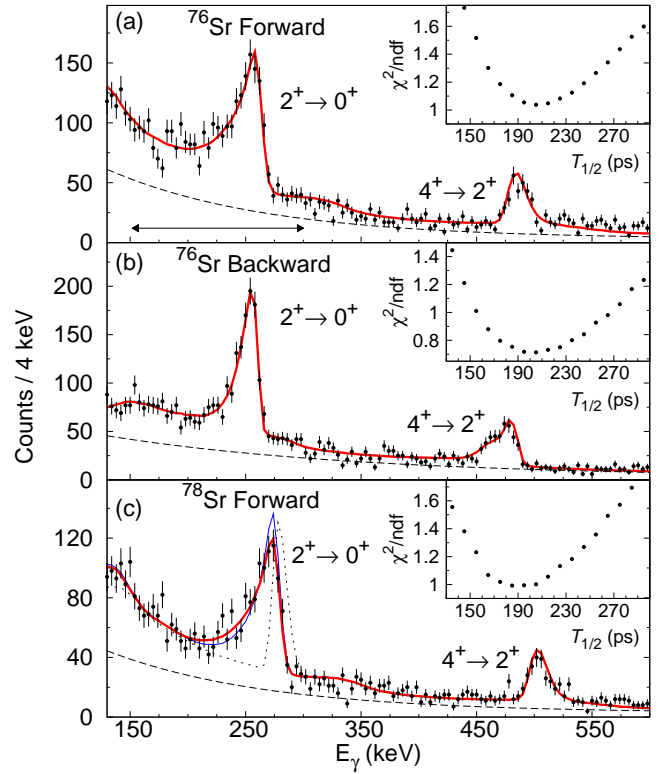}
  \caption{\label{fig:fig2}  (color  online)  Doppler-shift  corrected
    $\gamma$-ray spectra obtained for $^{76}$Sr in the (a) forward and
    (b) backward  rings, and  (c) for $^{78}$Sr  in the  forward ring.
    The data  are compared to  the simulated spectra (the  thick (red)
    curves)  which   include  background  contributions   (the  dashed
    curves).  The  insets show  the reduced $\chi^2$  distributions in
    the fit.   The range of data used  in the fit is  indicated by the
    arrow in  (a).  In (c), the  simulated spectra are  also shown for
    the  previous data  of  $T_{1/2}$~=~155~ps~\cite{Lis82} (the  thin
    (blue) solid curve) and a reference value of $T_{1/2}$~=~0~ps (the
    dotted curve).}

\end{center}
\end{figure}%
The outgoing particles  were identified (Fig.~\ref{fig:fig1}(a)) based
on   the  time-of-flight  and   energy-loss  measurements   using  the
focal-plane detection system of the S800 spectrometer~\cite{Baz03}.

De-excitation  $\gamma$ rays  were detected  by 15  Ge  detectors from
SeGA~\cite{Mue01}.
Each Ge crystal has a diameter of 7 cm and is divided into eight 1-cm
wide segments along the crystal length.
The   detectors  were  arranged   around  the   target  in   a  barrel
configuration with the  long side of the crystal  parallel to the beam
axis.
Two rings of  7 and 8 detectors were used to  cover the forward angles
of  50$-$80$^{\circ}$ and the  backward angles  of 95$-$125$^{\circ}$,
respectively.
The  full-energy peak efficiency  was measured  to be  17.5(3)~$\%$ at
244~keV by a standard $^{152}$Eu source.
The  present  setup  was  chosen to  maximize  $\gamma$-ray  detection
efficiencies as  well as  the sensitivity to  lifetime effects  on the
$\gamma$-ray line-shape as explained later.
Figure~\ref{fig:fig1}(b)  shows an  energy  spectrum of  $\gamma$~rays
measured  in  coincidence  with  $^{76}$Sr,  where  the  Doppler-shift
correction was made by assuming  that all $\gamma$~decays occur in the
middle    of    the   target    with    an    average   velocity    of
$\beta_{\text{mid}}$=$v/c=0.396$.
The $\gamma$-ray peaks are evident for the yrast band from the 2$^{+}$
to  the  8$^{+}$ states,  demonstrating  the  ability  of the  present
reaction to populate medium-spin states.
Inclusive  populations,  which are  the  sum  of  direct and  indirect
populations,   are   estimated  to   be   51(12)$\%$  and   36(8)$\%$,
respectively, for  the 2$^{+}$ and 4$^{+}$ states,  showing that about
70$\%$ of  the 2$^{+}$ state population  was made by  feeding from the
4$^{+}$ state.

In this work, the lifetimes of the 2$^{+}$ states of $^{76,78}$Sr were
determined  by the $\gamma$-ray  line-shape method~\cite{Ter08,Doo10},
which  is based on  the emission-point  distribution of  $\gamma$ rays
emitted from reaction residues in flight.
At the current beam  velocities of $v/c\approx0.4$, 
if an excited-state lifetime is on the order of 100~ps,
the $\gamma$ decay occurs, on average, about 1~cm behind the target.
Since we assume  the $\gamma$-ray decay occurs at  the target position
to   define  the  $\gamma$-ray   emission  angles   for  Doppler-shift
corrections, the  lifetime effect results  in a low-energy tail  for a
$\gamma$-ray peak as well as a slightly lower final peak position.
To  maximize the  sensitivity of  the $\gamma$-ray  line-shape  to the
lifetime, we produced Doppler-shift corrected spectra of $^{76,78}$Sr
by using velocities of outgoing Sr ions measured event-by-event in the
S800  (the averaged  velocities were  $\beta_{\text{aft}}=$  0.335 for
$^{76}$Sr and 0.330  for $^{78}$Sr).  
As  shown in  Fig.~\ref{fig:fig2}, asymmetric  shapes  of $\gamma$-ray
peaks are clearly  seen for the 2$^{+}$~$\rightarrow$~0$^+$ transition
both  in $^{76}$Sr  (Figs.~\ref{fig:fig2}~(a) and  (b))  and $^{78}$Sr
(Fig.~\ref{fig:fig2}(c)).
However, the 4$^{+}$ peaks are not aligned between the forward and
backward data which indicates that the 4$^{+}$ state decays mostly inside
the target  with higher recoil velocities. 
This suggests that  the lifetime of the 4$^{+}$  state is much shorter
than the flight time ($\approx$20~ps) of the ejectiles passing through
the target.

Lifetimes were obtained by comparing the measured spectra to simulated
ones as shown in Fig.~\ref{fig:fig2}.
The   simulation   is  based  on an  existing   code  which   utilizes
GEANT4~\cite{Adr09}, with modifications to incorporate the geometry of
the present setup.
The least $\chi^{2}$ method was used in a fitting procedure, where the
lifetime of the 2$^{+}$ state and amplitudes of the simulated spectrum
for the  2$^{+}$$\to$0$^{+}$ transition and  an exponential background
were taken as variable parameters.
The exponential slope of the  background was found to be
the  same for  other reaction products and  was thus  fixed  for both
$^{76,78}$Sr analyses.
The spectral shapes of $\gamma$  rays depopulating  
the 4$^+$, 6$^+$, and 8$^+$  states were included in the final fit,
where the amplitudes were determined in a different fit.
The lifetimes of the 4$^{+}$ states of $^{76,78}$Sr were both fixed to
be      equal      to      that     for      $^{78}$Sr      (half-life
$T_{1/2}$~=~5.1~ps)~\cite{Lis82}.
Based on the  reduced $\chi^{2}$ distributions as shown  in the insets
of Fig.~\ref{fig:fig2}, $T_{1/2}$ of  the 2$^+$ state in $^{76}$Sr was
found  to  be $207^{+16}_{-14}$~ps  and  $203^{+18}_{-16}$~ps for  the
forward and backward data, respectively.
For   $^{78}$Sr,    $T_{1/2}$~=~$188^{+17}_{-15}$~ps   (forward)   and
$194^{+21}_{-19}$~ps (backward) were obtained.
Systematic errors  were mainly due  to ambiguities in the  geometry of
the  setup~(3$\%$),   the  feeding  from  the   4$^+$  state  (1$\%$),
$\gamma$-ray anisotropy  effects~(1.5$\%$), and the  assumption of the
background~(3$\%$).
The overall systematic  error in the present measurement  was taken to
be 4.6$\%$ by adding these uncertainties in quadrature.

By combining the  forward and backward data, the  present results were
determined   to   be    $T_{1/2}$~=~205(25)~ps   for   $^{76}$Sr   and
$T_{1/2}$~=~191(27)~ps for  $^{78}$Sr, where both  the statistical and
systematic errors are included.
The present  result for $^{78}$Sr  is slightly larger,  but consistent
with the previous data of $155 (19)$~ps~\cite{Lis82}.
By  adopting $E$($2^+$)  = 262.3  keV for  $^{76}$Sr  \cite{Dav07} and
277.6~keV  for $^{78}$Sr~\cite{Rud97}, the  $B$(E2$\downarrow$) values
are  determined   to  be  2220(270)~e$^{2}$fm$^{4}$~for~$^{76}$Sr  and
1800(250)~e$^{2}$fm$^{4}$~for~$^{78}$Sr.
Note  that  main sources  of  the  systematic  errors are  common  for
$^{76,78}$Sr,  and   thus  the  present  results   indicate  that  the
collectivity of  $^{76}$Sr is larger  than that of $^{78}$Sr  by about
2$\sigma$.
Following the  prescription from Ref.~\cite{Ram01} for  a rigid rotor,
deformation  parameters  are  obtained  as  $\beta_2  =  0.45(3)$  for
$^{76}$Sr and $0.40(3)$ for $^{78}$Sr.

\begin{figure}
\begin{center}
  \includegraphics[width=1.\columnwidth]{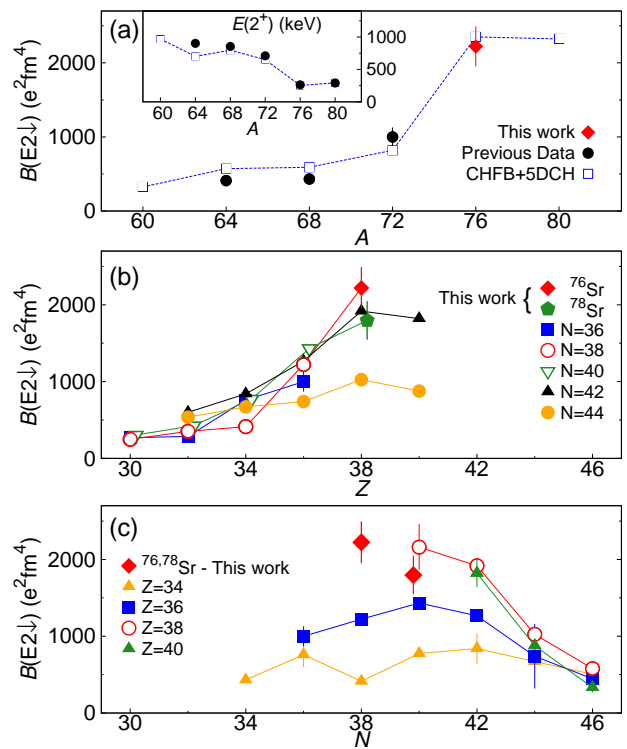}
  \caption{\label{fig:fig3} (color online)  (a) Experimental values of
    $B$(E$2\downarrow$)  and $E$(2$^{+}$)  (inset)  for the  $N$~=~$Z$
    nuclei   are   compared   to   predictions  from   the   CHFB+5DCH
    calculations~\cite{Del10}.  The  $B$(E$2\downarrow$) values in the
    vicinity of $^{76}$Sr are also plotted in the (b) isotonic and (c)
    isotopic       chains.        Data       are      taken       from
    \cite{Ram01,Lis82,Sta07,Gad05,Obe09,Gor05}.  }
\end{center}
\end{figure}

The  systematic behavior of  the $E$(2$^{+}$)  and $B$(E2$\downarrow$)
values  in  the  vicinity  of  $^{76}_{38}$Sr$_{38}$  are  plotted  in
Fig.~\ref{fig:fig3}.
Along the $N$~=~$Z$ line,  the $B$(E2$\downarrow$) data depict a rapid
increase of collectivity from $^{72}$Kr to $^{76}$Sr, accompanied by a
sudden decrease in $E$(2$^{+}$) (Fig.~\ref{fig:fig3}(a)).
This is  consistent with the occurrence  of the deformed  shell gap at
the  nucleon number  38 in  the Nilsson  diagram~\cite{Naz85}. 
However  this  scheme does  not  easily  account  for possible  mutual
effects  of proton  and neutron  deformation driving  contributions at
$N$~=~$Z$.
Such effects  are studied in  Figs.~\ref{fig:fig3} (b) and  (c), where
the $B$(E2$\downarrow$)  data are plotted  as a function of  $Z$ ($N$)
for the isotonic (isotopic) chain around $^{76}$Sr.
In Fig.~\ref{fig:fig3}(b), the  collectivity increases toward $Z$~=~38
for all the  isotonic chains, while the enhancement  is largest in the
$N$~=~38 chain.
For the isotopic  chains (Fig.~\ref{fig:fig3}(c)), the collectivity is
enhanced at $N$~=~38 only when $Z$~=~38.
This observation indicates  that the deformed shell gap  at the single
nucleon number 38 is not  strong enough to induce a large ground-state
deformation and a mutual support from proton and neutron contributions
is essential for the enhanced collectivity observed for $^{76}$Sr.

From      a      theoretical      point     of      view,      various
works~\cite{Naz85,Bon85,Lal95,Sar09,Pet02,Mar92,Lan03,Has07,Del10}
have attempted  to describe the ground-state deformation  of nuclei in
this  region.  
In  Fig.~\ref{fig:fig3}(a),  the  experimental  data of  $E$($2^+$)  and
$B$(E2)    are    compared     to    the    predictions    from    the
constrained-Hartree-Fock-Bogoliubov theory together  with a mapping to
the five-dimensional collective Hamiltonian (CHFB+5DCH)~\cite{Del10}.
Recently,  improvements to  mean-field  theories involving  quadrupole
correlations~\cite{Ben06}      and       mixing      of      different
deformations~\cite{Ben06,Del10}  have been  undertaken to  account for
the mutually enhanced magicity~\cite{Eld83}.  A good agreement between
the     present     data     and    the     CHFB+5DCH     calculations
(Fig.~\ref{fig:fig3}(a))   suggests  similar  improvements   are  also
required  to   account  for  the  evolution   of  collectivity  around
$^{76}$Sr.  Particularly, the CHFB+5DCH  theory takes into account the
mixing  of different shapes  including a  triaxial degree  of freedom,
reproducing remarkably well the trend and amplitude of the data in the
$A$~$\sim$~70 region~\cite{Obe09,Gir09} with pronounced prolate-oblate
shape coexistence.
While  predictions for  spectroscopic information  are  not available,
large  deformation for  ($^{76}$Sr, $^{78}$Sr)  are also  predicted by
other frameworks as  (0.37, 0.37)~\cite{Naz85}, (0.45, 0.45)~(RMF with
NL-SH interaction~\cite{Lal95}), (0.42, 0.42)~(FRDM~\cite{Mol95}), and
(0.44, 0.43)~(ETF-SI~\cite{Abo95}).

The microscopic origin of  the occurrence of the enhanced collectivity
in $A$~$\sim$~80 nuclei  is ascribed to the occupation  of nucleons in
the $g_{9/2}$ orbital~\cite{Naz85,Lan03,Has07}.
The  effect due to the $g_{9/2}$ intrusion
can  be  clearly   seen  in  the  sudden  increase  of
collectivity  from   $^{72}_{36}$Kr$_{36}$  to  $^{76}_{38}$Sr$_{38}$,
where  the occupation  numbers for  the  protons and  neutrons in  the
$g_{9/2}$  orbital   are  both   predicted  to  increase   from  about
2~to~3~\cite{Lan03}.
Interestingly, the  $g_{9/2}$ occupation  of neutrons is  predicted to
further increase from $N$~=~38  to 40 in the Sr isotopes~\cite{Lan03},
while the present results show that the maximum collectivity occurs in
$^{76}$Sr with $N$~=~$Z$~=~38.
This suggests  that the deformation  driving effects are  saturated at
the nucleon  number of 38  and hence one  would expect there to  be no
additional increase of collectivity in $^{80}$Zr.

\begin{figure}
\begin{center}
  \includegraphics[width=1.\columnwidth]{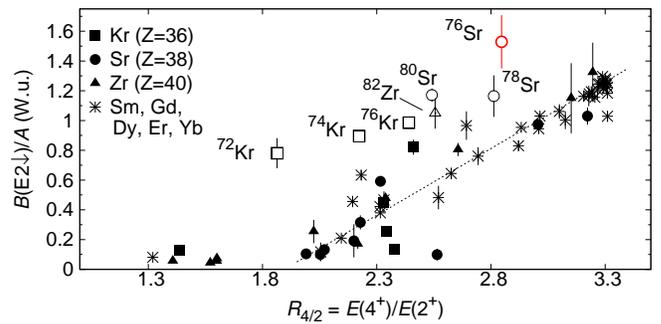}
  \caption{\label{fig:fig4}  (color   online)  Measured  $B$(E$2$)/$A$
    values          against         the          energy         ratios
    $R_{4/2}$~=~$E$(4$^{+}$)/$E$(2$^{+}$) are plotted  for the Kr, Sr,
    and  Zr isotopes  around $Z$~=~38  and the  $Z$~=~62$-$70 isotopes
    (Sm,   Gd,    Dy,   Er,   and   Yb).    Data    are   taken   from
    Ref.~\cite{Ram01,Gad05,Gor05} except for $^{76,78}$Sr where present data
    are shown.  The dashed line is to guide the eye.}
\end{center}
\end{figure}

To better characterize the  collective nature of $^{76}$Sr, a possible
signature of shape coexistence  phenomena can be investigated based on
the systematic  behavior for  the $B$(E2) with  respect to  the energy
ratio $R_{4/2}$.
If  two different configurations  coexist, the  mixing among  them can
lead to  a reduced $R_{4/2}$ of  the yrast band, as  the mixing lowers
the   ground    0$^{+}$   state   significantly    more   than   other
states~\cite{Cas93}.
In Fig.~\ref{fig:fig4}, we plot  the correlation between $R_{4/2}$ and
$B$(E2$\downarrow$)/$A$ for the present results of $^{76,78}$Sr.
For a systematic comparison, data  are also plotted for neighboring Kr
($Z$~=~36),  Sr ($Z$~=~38),  and  Zr ($Z$~=~40)  isotopes  as well  as
heavier mid-shell nuclei with $Z$~=~62$-$70.
As discussed in Ref.~\cite{Cas93}, the $B$(E2) values, when divided by
$A$,  have   a  clear   correlation  with  $R_{4/2}$,   starting  from
vibrational nuclei  with $R_{4/2}$ of  2.0 and evolving  to rotational
nuclei with $R_{4/2}$ of 3.3.
In fact, the correlation is evident in Fig.~\ref{fig:fig4} for many of
the Kr, Sr, and Zr isotopes,  shown by the closed symbols, and most of
the heavier nuclei.
However,   the  present   result  for   $^{76}$Sr,  as   well   as  the
$A$~=~70~$\sim$~80   nuclei  highlighted  by   the  open   symbols  in
Fig.~\ref{fig:fig4}, significantly deviates  from the global behavior,
suggesting the signature of shape coexistence.
A unique feature  for $^{76}$Sr is that the  mixing amplitude obtained
with a typical mixing strength of 0.2~MeV~\cite{Kor01} and a predicted
second  0$^{+}$ state  around 1~MeV~\cite{Has07,Del10}  is  very small
($\sim$5$\%$), preserving the large ground-state deformation.
However, the mixing  effect can be amplified in  the excitation energy
information  due  to  the  small $E$(2$^{+}$),  masking  the  deformed
character of $^{76}$Sr in $R_{4/2}$.
This emphasizes the importance of the present $B$(E2$\downarrow$) data
as a direct indicator of the enhanced collectivity of $^{76}$Sr.

In  summary,  the present  work  demonstrated  the  usefulness of  the
$\gamma$-ray line-shape method combined with two-step nucleon exchange
reactions  in  excited-state   lifetime  measurements,  extending  the
$B$(E2) systematics among self-conjugate nuclei up to $N$~=~$Z$~=~38.
The  results indicate  a large  ground-state deformation  of $^{76}$Sr
with $\beta_2$~=~0.45(3) despite the unusually low $R_{4/2}$ ratio and
illustrate the mutual enhancement of collectivity that uniquely occurs
at $N$~=~$Z$~=~38.
The comparison with theoretical  predictions as well as the systematic
behaviour  of   the  $R_{4/2}$  and  $B$(E2)   values  highlights  the
importance  of  the  mixing   of  coexisting  shapes  for  a  rigorous
description  of well-deformed  nuclei in  the  $A$~$\sim$~80 $N$~=~$Z$
region.
This  work  is supported  by  the  National  Science Foundation  under
PHY-0606007 and PHY-1102511 and by the UK STFC.

\bibliography{biblio}

\end{document}